\documentclass[a4paper]{article}
\usepackage[psamsfonts]{amssymb}
\usepackage{amsmath}
\usepackage{epsfig}

\author{H. Mohseni Sadjadi\footnote{mohseni@phymail.ut.ac.ir}
\\ {\small School of Physics, University College of Science,  University of Tehran,}
\\ {\small North Karegar Ave.  Tehran, Iran.}}
\title{ Generalized second law in phantom dominated universe}

\begin{document}
\maketitle
\begin{abstract}
We study the conditions of validity of the generalized second law
in phantom dominated era.\newline PACS: 98.80.-k, 98.80.Jk

\end{abstract}
\section{Introduction}
Astrophysical data show that the equation of state parameter
$\omega$ lies near $\omega=-1$ and possibly $\omega<-1$, leading
to an accelerated universe \cite{Han}. Some present data seem to
favor an evolving dark energy with $\omega$ less than $-1$ at
present epoch from $\omega
> -1$ in the near past \cite{Bo}. So we may assume that the
universe is dominated by a perfect fluid for which $\omega<-1$,
dubbed phantom energy \cite{cal, cal2}. This description of the
universe may contain finite-time future singularity, accompanied
with divergence of dark energy density, called Big Rip \cite{cal}.
Depending on the phantom potential, this model may also lead to
universes expanding for ever, e.g., universe tending to a de
Sitter space time \cite{fa}. The effects of gravitational
back-reactions can also counteract that of phantom energy and can
become large enough to terminate the phantom dominated phase
before the big rip \cite{wu}.

Thermodynamics of the expanding universe has also been the subject
of several studies
\cite{bru,li,gonz,dav1,dav2,bous,pav1,pav2,noj,Qing}. Phantom
thermodynamics looks leading to negative entropy of the universe
\cite{brev} or to appearance of negative temperatures \cite{sig}.
In accelerated expanding universe, besides the normal entropy, a
cosmological horizon entropy can also be considered. One can
investigate the conditions for which the generalized second law of
thermodynamics (GSL) holds \cite{dav1, dav2}. In these cases GSL
asserts that the sum of the horizon entropy, and the normal
entropy of the fluid is an increasing function of time. In
\cite{dav1} the change in event-horizon area in cosmological
models that depart slightly from de Sitter space was investigated,
and it was shown that the area and consequently the (de Sitter)
horizon entropy are non decreasing functions of time. In the
presence of a viscous fluid, there was found that GSL was
satisfied provided that the temperature of the fluid was equal to
or lower than de Sitter horizon temperature, $T_{dS}=H/(2\pi)$,
where $H$ is the Hubble parameter. In \cite{pav1} the GSL for two
specific examples of phantom dominated universe was discussed. In
the first example it was shown that for a time independent
parameter of state, the total entropy is a constant. In the second
example using a power law potential and the slow climb
approximation, it was shown that GSL was satisfied. In the absence
of a well defined Hawking temperature for cosmological horizon,
the temperature of the phantom fluid in \cite{pav1} was assumed to
be the same as the de Sitter temperature.

The phantom universe, has a future event horizon (cosmological
horizon), $R_h$. In this model the Hubble parameter is an
increasing function of time so we expect that the horizon entropy,
in contrast to the model discussed in \cite{dav1}, decreases with
time.

In this paper we consider the phantom dominated era, and prove
that the future event horizon area is a non-increasing function of
time. It is shown that one necessary condition for satisfying GSL
is the positivity of the temperature. If the temperature is
assumed to be proportional to the de Sitter temperature
\cite{dav1}
\begin{equation}\label{1}
T=\frac{bH}{2\pi},
\end{equation}
an inequality in terms of Hubble parameter, $H$, the future event
horizon, $R_h$, and the parameter $b$, will be obtained. By
solving this inequality one can determine the dynamics of $R_h$,
$H$, and also the range of $b$. We show that GSL is satisfied for
a wide range of models, provided that $0<b\leq 1$. We also study
the condition of validity of GSL in the transition from
quintessence regime to the phantom dominated era. At the end, two
specific examples are used to emphasize our general results.

We use the units $\hbar=c=G=k_{B}=1$.

\section{Phantom thermodynamics}
The spatially flat FRW  universe in comoving coordinates
$(t,x,y,z)$, is described by the metric
\begin{equation}\label{2}
ds^2=-dt^2+a^2(t)(dx^2+dy^2+dz^2).
\end{equation}
The relative expansion velocity of the universe, in terms of the
scale factor $a(t)$, is given by the Hubble parameter
$H=\dot{a}/a$. The over dot indicates derivative with respect to
the comoving time $t$. We assume that the universe is filled with
a perfect fluid with the energy density $\rho$, and the pressure
$P$. The fluid state parameter $\omega$ is defined through the
equation of state
 $P=\omega \rho$. Einstein equations are
\begin{eqnarray}\label{3}
{dH\over{dt}}&=&-4\pi(P+\rho) \nonumber \\
H^2&=&{8\pi\over 3}\rho.
\end{eqnarray}
The above equations result
\begin{equation}\label{5}
\omega=-1-{2\dot{H}\over {3H^2}}. \end{equation} For an
accelerating universe we have $\dot{H}+H^2>0$, which, in terms of
$\omega$, can be expressed as $\omega<-{1\over 3}$. The
corresponding cosmological fluid is then consisted of some sort of
energy density which has a negative pressure known as dark energy.
Matter with $\omega<-1$ is dubbed phantom energy. By considering
(\ref{5}), one can see that for phantom fluid: $\dot{H}
>0$, which using (\ref{3}) results
$P+\rho<0$.

There are various approaches to study the dark energy. One
approach is introducing scalar fields. The behavior of $\omega$ in
the phantom regime, $\omega<-1$, can be related to the presence of
a phantom scalar field, $\phi$, with a wrong sign kinetic term
\cite{cal}. Depending on the form of the phantom scalar field
potential, different solutions such as asymptotic de Sitter, big
rip, and so on may be obtained \cite{fa}. To describe $\omega>-1$,
or quintessence regime, a normal scalar field, $\sigma$, known as
quintessence scalar field, can be used \cite{pe}. The phase
transitions in inflationary models can be investigated using
hybrid models. These models, at least, are composed of two scalar
fields \cite{linde}, e.g., one of these models is the quintom
model which in order to describe the transition from $\omega>-1$
to $\omega<-1$ regime, assumes that the cosmological fluid,
besides the ordinary matter and radiation, is consisted of a
quintessence and a phantom scalar field \cite{quin}.

Another method to study the acceleration of the universe is to use
a running cosmological constant based on the principles of quantum
field theory (specially on the renormalization group) which can
mimic the behavior expected for quintessence and phantom -like
representations of the dark energy. In this method despite of the
absence of scalar fields, an effective equation state like
(\ref{5}) can be obtained \cite{sola}.

In this paper we will use the equations (\ref{3}), (\ref{5}), and
assume that the universe is filled with a perfect fluid in phantom
phase in the sense that $\omega<-1$, but our results will be
independent of the scalar fields or other origins of the phantom
energy.

Consider the era when phantom phase is dominated i.e, $\dot{H}>0$.
We assume that $a(t)\rightarrow \infty$ when $t\rightarrow t_s$,
so the scale factor will diverge for a future value on the world
time: $t_s$. The radius of the observer's (future) event horizon
is
\begin{equation}\label{6}
R_h=a(t)\int_t^{t_s}{dt'\over{a(t')}},\qquad
\int_t^{t_s}{dt'\over{a(t')}}<\infty.
\end{equation}
Using (\ref{6})one can see that $R_h$ satisfies the following
equations
\begin{eqnarray}\label{7}
\dot{R_h}&=&HR_h-1, \nonumber \\
 \ddot{R_h}&=&(\dot{H}+H^2)R_h-H.
\end{eqnarray}
For de Sitter space time, $t_s=\infty$, $R_h=1/H$, and the
eq.(\ref{7}) continues to hold.

{\it{Theorem: $R_h$ is a non-increasing function of time,
$\dot{R_h}\leq 0$}}.\newline {\it{Proof}}: We have $\dot{H}\geq
0$, therefore $\ddot{a}\geq {\dot{a}^2/a}$. This inequality can be
integrated from $t$ to $t_s$
\begin{equation}\label{8}
\int^{t_s}_t {\ddot{a}(t')\over {\dot{a}}^2(t')}dt'\geq
\int^{t_s}_t {1\over a(t')}dt',
\end{equation}
resulting
\begin{equation}\label{9}
{1\over{\dot{a}}(t)}-{1\over{\dot{a}}(t_s)}\geq \int^{t_s}_t
{1\over a(t')}dt'.
\end{equation}
Suppose that ${1\over{\dot{a}}(t_s)}$ is bounded from below by a
positive number $\epsilon$. In our (super)accelerating model of
universe, $\ddot{a}>0$ yields
\begin{equation}\label{10}
{1\over{\dot{a}(t)}}>{1\over{\dot{a}}(t_s)}>\epsilon, \qquad
t<t_s.
\end{equation}
Therefore
\begin{equation}\label{11}
\int^{\infty}_a{da\over{\dot{a}a}}>\epsilon
\int^{\infty}_a{da\over{a}}=\infty.
\end{equation}
But (\ref{11}) conflicts with the assumption of having a future
horizon(\ref{6}). Therefore $1/{\dot{a}(t_s)}=0$, and using
(\ref{9}), we obtain $HR_h\leq 1$. Using this result and
(\ref{7}), we arrive at $\dot{R_h}\leq 0$.

The future event horizon area is $4\pi R_h^2$. We assume that this
horizon area represents a true contribution to the total entropy.
The corresponding entropy is proposed to be \cite{pav1}
\begin{equation}\label{12}
S_h=\pi R_h^2.
\end{equation}
In this way $\dot{S_h}=2\pi R_h \dot{R_h}$ which is negative and
shows that $S_h$ decreases with time. This is in contrast to the
assumption of \cite{gong}, upon which it was argued that the
holographic dark energy has no phantom-like behavior.

The entropy of the phantom fluid inside the cosmological horizon
of a comoving observer is related to the energy and the pressure
via the first law of thermodynamics
\begin{equation}\label{13}
TdS=dE+PdV =(P+\rho)dV+Vd\rho.
\end{equation}
In terms of $H$ this law can be written as
\begin{equation}\label{14}
TdS=-{1\over {4\pi}}{dH\over {dt}}dV+Vd\rho.
\end{equation}
Using $V={4\over 3}\pi R_h^3$ and $H^2=(8\pi/3)\rho$, we obtain
\begin{equation}\label{15}
TdS=-\dot{H}R_h^2dR_h+HR_h^3dH.
\end{equation}
Then using (\ref{7}), we arrive at
\begin{equation}\label{16}
T\dot{S}=\dot{H}R_h^2.
\end{equation}
For phantom fluid $\dot{H}>0$, $\dot{S}>0$, provided we take
$T>0$. GSL states that the sum of the ordinary entropy and the
horizon entropy cannot decrease with time
\begin{equation}\label{17}
\dot{S}+\dot{S_h}\geq 0.
\end{equation}
In our model this law yields
\begin{equation}\label{18}
{\dot{H}\over T}R_h+2\pi \dot{R_h}\geq 0.
\end{equation}

Note that in phantom regime, $\dot{H}>0$ and $\dot{R_h}<0$, so, a
necessary condition that the GSL to be satisfied is positivity of
the phantom fluid temperature $T>0$. Therefore the GSL is violated
in phantom models with positive entropy and negative temperature
\cite{sig}. For a discussion about this subject see \cite{noj1},
\cite{pav1}.

The GSL asserts that the entropy of an isolated system cannot
decrease, but does not determine the sign of the entropy. But if
we assume that at the finite time $t=t_s$(i.e. when $R_h=0$) we
have $S=0$ \cite{pav1}, and the GSL is assumed to be satisfied in
an era before $t_s$, then the entropy will be negative in this
epoch, which agrees to \cite {brev} and \cite{noj1} in which a
negative entropy for a specific model of phantom thermodynamics
with positive temperature has been derived. However in the above
mentioned papers i.e., \cite{brev}, \cite{noj1} and \cite{sig},
the role of the future event horizon has been ignored.

If the temperature is taken to be $T={bH\over {2\pi}}$, GSL
results
\begin{equation}\label{19}
b\leq -{\dot{H}\over H}{R_h\over \dot{R_h}}.
\end{equation}
For a de Sitter space-time $R_h={1\over H}$, which results $b\leq
1$, so we expect that in a phantom model which is small perturbed
around de sitter space the fluid temperature is equal to or less
than the de Sitter temperature.

If the total entropy is increasing in the universe expansion,
\begin{equation}\label{20}
\dot{S}+\dot{S_h}> 0,
\end{equation}
equation (\ref{19}) results
\begin{equation}\label{21}
(H^{1\over b}R_h\dot{)}>0.
\end{equation}
For $b=1$, we must have $(HR_h\dot{)}>0$, or using (\ref{7})
\begin{equation}\label{22}
\ddot{R_h}>0.
\end{equation}
Here $\dot{R_h}$ is an increasing negative function of time. When
$\dot{R_h}\rightarrow 0^{-}$,  $R_h$ tends to the hubble radius
$1/H$ (see equation (\ref{7})).

\begin{figure}
\centering\epsfig{file=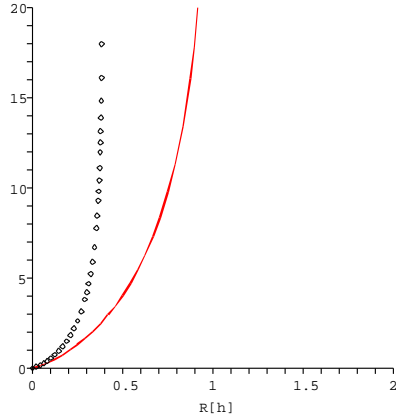,width=6cm} \caption{$t_s-t$ as a
function of $R_h$, for $b=0.9$, $\lambda=1$ (continuous line) and
$\lambda=1.1$ (points).} \label{fig1}
\end{figure}

Now let us assume that the expansion of the universe is a
reversible adiabatic process such that
\begin{equation}\label{23}
\dot{S}+\dot{S_h}= 0.
\end{equation}
From (\ref{19}) we obtain $HR_h^b=\lambda$ where $\lambda$ is a
constant. \newline For $b=1$, using (\ref{7}), this results
$\dot{R_h}=\lambda-1$ and using the fact that $R_h(t_s)=0$, for
$\lambda \neq 1$, we obtain
\begin{equation}\label{24}
R_h=(1-\lambda)(t_s-t).
\end{equation}
Note that $\dot{R_h}<0$, therefore $\lambda< 1$. $\omega$ is given
by
\begin{equation}\label{25}
\omega=-{1\over 3}-{2\over 3\lambda}<-1.
\end{equation}
If we take $t_s=\infty$, $R_h$ becomes infinite as claimed in
\cite{gonz}. If $\lambda=1$, the space-time is de Sitter,
$R_h={1\over H}$ and $\omega=-1$.\newline For $b\neq 1$,
(\ref{23}) and the first equation of (\ref{7}) lead to the
following equation
\begin{equation}\label{26}
\dot{R_h}-\lambda R_h^{1-b}+1=0,
\end{equation}
with solutions satisfying
\begin{equation}\label{27}
{{R_h\Phi(\lambda R_h^{1-b},1,{1\over{1-b}})}\over {1-b}}=t_s-t.
\end{equation}
$\Phi$, is the Lerchiphi function. The condition $\dot{R_h}<0$
yields $\lambda<R_h^{b-1}$, but $R_h(t_s)=0$, so to have
$\lambda\neq 0$, $b$ must satisfy the inequality $b<1$, i.e,
$T<{H\over {2\pi}}$. $\dot{R_h}=0$ occurs at
$R_h=\lambda^{1\over{b-1}}$, which is a branch point of $\Phi$. At
$t=t_s$ we have $R_h=0$ as expected. See fig. (\ref{fig1}).

The above results are general and do not depend on the previous
phase of the phantom universe. As it was mentioned in the
introduction, based on recent astrophysical data, one can consider
the possibility of the transition from $\omega>-1$ to $\omega<-1$.
So it may be interesting to see how our results may be restricted,
if the GSL still hold in the transition epoch. By transition epoch
we mean a little neighborhood around the transition time: $t_t$.
Like the phantom universe, for a quintessence universe we can
consider the cosmological event horizon via (\ref{6}). The
relation (\ref{18}) sill valid. By considering that $\dot{H}<0$,
we deduce that a necessary condition for satisfying GSL in such
universe is $\dot{R_h}>0$. But in the phantom universe, $R_h$ is
non-increasing. Therefore for a continuous $\dot{R_h}$ (which may
be the result of continuity of $H$ and $R_h$, see (\ref{7})), we
must have $\dot{R_h}(t_t)=0$.  Using (\ref{7}), this results
$HR_h=1$ at $t_t$. At the transition time, we have also
$\dot{H}(t_t)=0$. In this way using (\ref{12}) and (\ref{16}), we
obtain $\dot{S}(t_t)=\dot{S_h}(t_t)=0$, therefore the total
entropy is differentiable and thereby continuous at $t_t$. For
$t<(>)t_t$, $S_h$ is an increasing (decreasing) function, while
the fluid entropy is a decreasing (increasing) function of time.
If the entropy is taken to be negative in the beginning of phantom
dominated era (which is not a consequence of GSL which considers
only the time derivative of the entropy) and the GSL is assumed to
be valid in the transition epoch, then we must have a negative
entropy in the quintessence era too.

\section{Examples}
To illustrate our general results, let us consider two examples.
As a first example consider the phantom universe of pole-like type
\begin{equation}\label{28}
a(t)=a_0(t_s-t)^{-n},
\end{equation}
$t\leq t_s$ and $n>0$ is a positive real number. $a_0$ is a real
positive constant. $\dot{H}>0$, and $\omega=-1-2/(3n)<-1$ is a
time independent constant. In this model $HR_{h}=n/(1+n)$, and
$\dot{R_h}=-1/(n+1)<0$. This is consistent with the theorem
imposed after the eq.(\ref{7}). Following the discussion after the
eq.(\ref{23}), and by taking $\lambda=n/(n+1)$, this model
corresponds to a reversible adiabatic expansion of an universe
with the phantom temperature $T=H/(2\pi)$ in agreement with the
results of \cite{pav1}.

To show the role of the parameter $b$ in validity of GSL, as an
another example, consider the model \cite{noj}
\begin{equation}\label{29}
a=a_0({t\over{t_s-t}})^n,
\end{equation}
where $a_0$ and $n$ are two positive real constants. In this model
\begin{eqnarray}\label{30}
H&=&n{t_s\over t(t_s-t)} \nonumber \\
\dot{H}&=&{nt_s(2t-t_s)\over{t^2(t_s-t)^2}}.
\end{eqnarray}
The condition $\dot{H}> 0$ is satisfied when $t_s<2t$. $\omega$ is
time dependent
\begin{equation}\label{31}
\omega=-1-{2(2t-t_s)\over {3nt_s}}.
\end{equation}
When $2t>t_s$, i.e., in phantom dominated era, we have
$\omega<-1$.

In this model the future event horizon is
\begin{equation}\label{32}
R_h=t_s(x-1)^{-n}\int^x_1(u-1)^nu^{-2}du,
\end{equation}
where $x:={t_s/ t}$, $1<x<2$. We have
\begin{equation}\label{33}
\dot{R_h}=-1+{nx^2(x-1)^{-n-1}}\int_1^x(u-1)^nu^{-2}du.
\end{equation}
Note that from the above equation one can show that $\dot{R_h}<0$,
in agreement with our previous results. This can be verified by
considering that for $x<2$ we have
\begin{equation}\label{34}
n\int_1^x(u-1)^nu^{-2}du<(\int_1^x{(u-1)^n\big((n-1)u+2\big)\over
u^3}du={(x-1)^{n+1}\over x^2}),
\end{equation}
Which can be verified by noting that the integrands satisfy the
above inequality for all $u$ belonging to $[1,2]$.

If GSL is respected, we must have $(H^cR_h{\dot )}>0$, where
$c:=1/b$. This condition can be written as
\begin{equation}\label{35}
(x-1)^{-n-1}\big((n-c)x^2+2cx\big)\int_1^x(u-1)^nu^{-2}du
>1.
\end{equation}
For  $1<x<2$, $n>0$, and $c>0$ (which follows from the discussion
after the eq.(\ref{18})), we have $(n-c)x^2+2cx>0$, therefore,
(\ref{35}) reduces to
\begin{equation}\label{36}
\int_1^x(u-1)^nu^{-2}du>{(x-1)^{n+1}\over{(n-c)x^2+2cx}}.
\end{equation}
The right side of the above inequality can be written as a
definite integral
\begin{equation}\label{37}
\int_1^x(u-1)^nu^{-2}du>\int^x_1(u-1)^nu^{-2}\Big[{(1-n)(c-n)u^2+2(n+cn-c)u+2c\over
\big( (c-n)u-2c\big)^2} \Big]du,
\end{equation}
which can be rewritten as
\begin{equation}\label{38}
\int^x_1{(u-1)^nu^{-2}\over{\big((c-n)u-2c\big)}^2}q du>0.
\end{equation}
$q$ is defined through
\begin{equation}\label{39}
q:=Au^2+2Bu+D,
\end{equation}
$A:=(c-1)(c-n)$, $B:=\big(n(c-1)+c-2c^2\big)$ and $D:=4c^2-2c$. If
for all $u$ in the interval $u\in[1,x]\subseteq [1,2]$, $q>(<)0$,
then the inequality (\ref{34}) holds (does not hold). At $u=1$,
$q(u=1)=(c-1)(n+c)$ and at $u=2$, $q$ is negative: $q(u=2)=-2c$.
Therefore a necessary condition for validity of GSL is that $q$
must have two roots such that at least one of them lies in
$[1,2]$. Otherwise $q$ will be negative in this interval.

If $n$ and $c$ satisfy
\begin{equation}\label{40}
n^2>{c^2(1-2c)\over{(c-1)^2}},
\end{equation}
$q$ has two roots.

For $A>0$, $q$ has a root $1<r<2$ at $r=(-B-\sqrt{B^2-AD})/A$,
provided that $c>1$. The other root takes place at
$r'=(-B+\sqrt{B^2-AD})/A>2$.

For $A<0$, $q$ has a root at $1<r=(-B-\sqrt{B^2-AD})/A<2$,
provided that $c>1$. The other root takes place at
$r'=(-B+\sqrt{B^2-AD})/A<1$.

In these cases we have $q(u=1)>0$, hence $q$ is positive for all
values of $u$ satisfying $1<u<r$ and negative for all values of
$u$ in the interval $r<u<2$. For $c<1$, neither of the roots of
$q$ lies in the interval$[1,2]$ and the sign of $q$ is negative.

For $A=0$ (in this case $q(u)$ is a line),  we have either $c=1$
or $c=n$. For $c=1$, $q(u=1)=0$ and $q$ is negative in $(1,2]$.
When $c=n$, we have $q(u=1)=2c(c-1)$ and $q(u=2)=-2c$, so if
$c<1$, $q$ will be negative in $[1,2]$.

Using these results we conclude that $q$ has a root in $[1,2]$,
provided that $c>1$. Other values of $c$ lead to negative values
of $q$ for all $u\in [1,2]$. Therefore $T<H/(2\pi)$ is a necessary
condition for validity of GSL.

\begin{figure}
\centering\epsfig{file=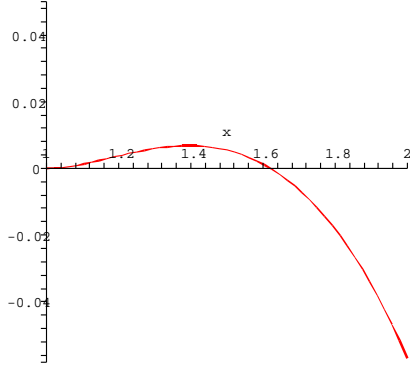,width=6cm}
\caption{$\int_1^x(u-1)^nu^{-2}du-(x-1)^{n+1}((n-c)x^2+2cx)^{-1}$,
as a function of $x$ for $n=1$ and $c=2$.} \label{fig2}
\end{figure}

Despite the above result, $c>1$ is not an enough condition for
validity of GSL, specially in the early period in which the
phantom became dominated as is illustrated in fig.(\ref{fig2}).
Using our arguments we can only claim that in the period $1<x<r$,
(in an interval before the big rip)the GSL holds. But note that
for large $c$ or large $n$ we have $r\simeq 2$, and  GSL is
applicable approximately in the whole region of the interval
$[1,2]$.

The transition from $\dot{H}<0$ to $\dot{H}>0$, occurs at $x=2$,
or $t={t_s\over 2}$, where $\dot{H}=0$, which results
$\dot{S}(t_t)=0$.  As we have discussed in the last part of the
previous section, a necessary condition for satisfying GSL in
$t<{t_s\over 2}$ is $\dot{R_h}(x=2)=0$ which using (\ref{33}),
implies
\begin{equation}\label{41}
4n\int_1^2(u-1)^nu^{-2}du=1.
\end{equation}
The above integral can be written in terms of digamma functions,
resulting
\begin{equation}\label{42}
4n\left(-{n\over 2(n+1)}-{1\over 2( n+1)}+{1\over 2} \,n \left(
\Psi \left( {1\over 2}+{1\over 2}\,n \right) -\Psi \left( {1\over
2}\,n \right)
 \right)  \right)=1.
\end{equation}
Following our previous discussion, we expect that GSL remains
valid in the region near $x=2$ for large $n$. By asymptotic
expanding of digamma functions, the above equation, (\ref{42}),
becomes
\begin{equation}\label{43}
1-{1\over{2n^2}}+O({1\over{n^4}})=1
\end{equation}
which is valid for large $n$. Therefore for large $n$, the GSL may
be true in the transition epoch. The condition (\ref{38}), in a
short time before the transition: $t=t_t-\epsilon=t_s/2-\epsilon$,
becomes
\begin{equation}\label{44}
\int^2_1{(u-1)^nu^{-2}\over{\big((c-n)u-2c\big)}^2}q du-{c\over
 8n^2}\epsilon>0,
\end{equation}
which based on our previous discussion about the enough condition
for validity of GSL, is correct for large $n$. Hence the GSL is
respected in the transition epoch only for large values of $n$.

\section{Summary}
In this paper we discussed the constraints and conditions imposed
on cosmological future horizon  $R_h$, Hubble parameter $H$, and
the temperature $T$, in a phantom dominated universe in order to
satisfy the GSL.  We proved that the future event horizon is a
non-increasing function(see the theorem after eq.(\ref{7})) and
thereby the corresponding horizon entropy is also a non-increasing
function of comoving time. To obtain this result we assumed that
one can relate an entropy to the future cosmological horizon, in
the same way as black-hole event horizons, despite their different
structures \cite{dav1,dav2,pav1}. Using the assumption that the
(super)accelerated universe is filled with a perfect fluid
satisfying the first law of thermodynamics (eq. (\ref{13})), we
obtained the total entropy and found that the positivity of the
temperature is a necessary condition for GSL (see eq.(\ref{18})).
We obtained also an inequality in terms of $R_h$ , $H$, and $T$
(see eqs.(\ref{19}), (\ref{21}) and (\ref{22})). This inequality
becomes an equation for the adiabatic reversible expansion of the
(super) accelerated universe. We solved this equation for $R_h$,
for de Sitter and non-de Sitter temperatures (see eqs.(\ref{24})
and (\ref{27})). We showed that in various cases $T$ must be less
than or equal to de Sitter temperature.  By studying the influence
of the transition from the quintessence to phantom dominated
universe on the GSL , we deduced that the time derivative of the
future event horizon and the entropy must be zero at the
transition time. However if we adopt that the GSL is valid in the
transition epoch and also assume that the entropy is negative in
the beginning of the phantom era (which is not a consequence of
GSL), a negative entropy in the quintessence regime will be
deduced, which is not expected.


\begin{thebibliography}{99}
\bibitem{Han}S. Hannestad and E. Mortsell, Phys. Rev. D {\bf 66}, 063508 (2002);
H. Jassal, J. Bagla and T. Padmanabhan, Phys.Rev. D {\bf 72},
103503 (2005); A. G. Riess et al., Astrophys. J. {\bf 607}, 665
(2004); U. Seljak et al., Phys. Rev. D {\bf 71}, 103515 (2005); D.
Huterer and A. Cooray, Phys. Rev. D {\bf 71}, 023506 (2005); Y.
Wang and M. Tegmark, Phys. Rev. D {\bf 71}, 103513 (2005); U.
Alam, V. Sahni and A. A. Starobinsky, JCAP {\bf 0406}, 008 (2004).
\bibitem{Bo}B. Feng, X. Wang and X. Zhang, Phys.Lett. B {\bf 607}, 35 (2005).
\bibitem{cal} R. R. Caldwell, Phys. Lett. B {\bf 545}, 23 (2002).
\bibitem{cal2} R. R. Caldwell, M. Kamionkowski and N. N. Weinberg,
Phys. Rev. Lett {\bf 91}, 071301 (2003).
\bibitem{fa} V. Faraoni, Class. Quant. Grav. {\bf 22}, 3235
(2005).
\bibitem{wu} P. Wu and H. Yu,  Nucl.Phys. B {\bf 727}, 355 (2005).
\bibitem{bru} R. Brustein, Phys. Rev. Lett. {\bf 84}, 2072
(2000).
\bibitem{li} M. Li, Phys. lett. B {\bf 603}, 1 (2004).
\bibitem{gonz} P. F. Gonzalez-Diaz, hep-th/0411070.
\bibitem{dav1} P. C. W. Davies, Class. Quant. Grav {\bf 4}, L225 (1987).
\bibitem{dav2} P. C. W. Davies, Class. Quant. Grav {\bf 5}, 1349 (1988).
\bibitem{bous} R. Bousso,  Phys. Rev.  D {\bf 71}, 064024 (2005).
\bibitem{pav1} G. Izquierdo and D. Pavon, astro-ph/0505601, to be appeared in Phys. Lett. B.
\bibitem{pav2} G. Izquierdo and D. Pavon, Phys. Rev. D {\bf 70}, 127505 (2004).
\bibitem{noj}  S. Nojiri and S. D. Odinstov, hep-th/0506212.
\bibitem{Qing} Q. Huang and M. Li,  JCAP {\bf 0408}, 013 (2004).
\bibitem{brev} I. Brevik, S. Nojiri, S.D. Odintsov and L. Vanzo,
Phys. Rev. D{\bf 70}, 043520(2004).
\bibitem{sig} P. F. Gonzalez-Diaz and C. L. Siguenza, Nucl. Phys. B {\bf 697}, 363 (2004);
E. Babichev, V. Dokuchaev and Y. Eroshenko, Phys. Rev. Lett. {\bf
93}, 021102 (2004).
\bibitem{pe} P. J. E. Peebles and A. Vilenkin, Phys. Rev.  D {\bf 59}, 063505 (1999).
\bibitem{linde} A. D. Linde, Phys. Rev. D {\bf 49}, 748 (1994) [astro-ph/9307002];
A. D. Linde, Phys. Lett. B {\bf 259}, 38 (1991); H. Wei and R.
Cai, astro-ph/0512018.
\bibitem{quin} B. Feng, X. L. Wang and X. M. Zhang, Phys. Lett. B {\bf 607}, 35 (2005);
 Z. K. Guo, Y. S. Piao, X. M. Zhang and Y. Z. Zhang, Phys. Lett. B {\bf 608}, 177 (2005);
 X. F. Zhang, H. Li, Y. S. Piao and X. M. Zhang, astro-ph/0501652.
\bibitem{sola}J. Sola and H. Stafinc, astro-ph/0507110; Phys. Lett. B {\bf 624} (2005) 147.
\bibitem{gong} Y. Gong, B. Wang and Y. Zhang, Phys. Rev. D {\bf 72}, 043510
(2005).
\bibitem{noj1} S. Nojiri, S. D.Odintsov, Phys. Rev. D {\bf 70}, 103522 (2004).

\end{thebibliography}
\end{document}